%%%%%%%%%1%%%%%%%%%2%%%%%%%%%3%%%%%%%%%4%%%%%%%%%5%%%%%%%%%6%%%%%%%%%7%%%%%%%%%
%%%%%%%%%%%%%%%%%%  tex macros for preprints, cm version %%%%%%%%%%%%%%
%                     (P. Ginsparg, last updated 9/91)
%                if confused, type `b' in response to query 
%
%---------------------------------------------------------------------%
%% site dependent options: 
%% \unredoffs and \redoffs define horizontal and vertical offsets 
%% respectively for unreduced and reduced modes. \speclscape defines
%% the \special{} call that sets printer to landscape (sideways) mode.
%% from standard set below, leave uncommented as appropriate or redefine
%
%%% next 400dpi
\def\unredoffs{} \def\redoffs{\voffset=-.31truein\hoffset=-.48truein}
\def\speclscape{}
%
%%% apple lw
%\def\unredoffs{} \def\redoffs{\voffset=-.31truein\hoffset=-.59truein}
%\def\speclscape{\special{ps: landscape}}
%
%%% qms lasergrafix:
%\def\unredoffs{} \def\redoffs{\voffset=-.4truein\hoffset=.125truein}
%\def\speclscape{\special{qms: landscape}}
%
%%% saclay A4 paper:
%\def\unredoffs{\hoffset-.14truein\voffset-.2truein} 
%\def\redoffs{\voffset=-.45truein\hoffset=-.21truein} 
%\def\speclscape{\special{landscape}}
%
%---------------------------------------------------------------------%
%
\newbox\leftpage \newdimen\fullhsize \newdimen\hstitle \newdimen\hsbody
\tolerance=1000\hfuzz=2pt
\catcode`\@=11 % This allows us to modify PLAIN macros.
\def\bigans{b }
%\message{ big or little (b/l)? }\read-1 to\answ
\def\answ{b }
\ifx\answ\bigans\message{(This will come out unreduced.}
%\magnification=1000
\unredoffs\baselineskip=16pt plus 2pt minus 1pt
\hsbody=\hsize \hstitle=\hsize %take default values for unreduced format
\else\message{(This will be reduced.} \let\l@r=L
\magnification=1000
%xxxxx
\baselineskip=16pt plus 2pt minus 1pt
%\baselineskip=12pt plus 2pt minus 1pt 
%xxxxx
\vsize=7truein
\redoffs \hstitle=8truein\hsbody=4.75truein\fullhsize=10truein\hsize=\hsbody
\output={\ifnum\pageno=0 %%% This is the HUTP version
  \shipout\vbox{\speclscape{\hsize\fullhsize\makeheadline}
    \hbox to \fullhsize{\hfill\pagebody\hfill}}\advancepageno
  \else
  \almostshipout{\leftline{\vbox{\pagebody\makefootline}}}\advancepageno 
  \fi}
\def\almostshipout#1{\if L\l@r \count1=1 \message{[\the\count0.\the\count1]}
      \global\setbox\leftpage=#1 \global\let\l@r=R
 \else \count1=2
  \shipout\vbox{\speclscape{\hsize\fullhsize\makeheadline}
      \hbox to\fullhsize{\box\leftpage\hfil#1}}  \global\let\l@r=L\fi}
\fi
%---------------------------------------------------------------------
%
\newcount\yearltd\yearltd=\year\advance\yearltd by -1900

\def\Title#1#2{\nopagenumbers\abstractfont\hsize=\hstitle\rightline{#1}%
\vskip 1in\centerline{\titlefont #2}\abstractfont\vskip .5in\pageno=0}
\def\Date#1{\vfill\leftline{#1}\tenpoint\supereject\global\hsize=\hsbody%
\footline={\hss\tenrm\folio\hss}}%      restores pagenumbers
%
%       use following instead of \Date on the preliminary draft, 
%       puts date/time on each page in big mode, writes labels in margins

\def\draftmode{\message{ DRAFTMODE }\def\draftdate{{\rm preliminary draft:
\number\month/\number\day/\number\yearltd\ \ \hourmin}}%
\headline={\hfil\draftdate}\writelabels
%xxx
\baselineskip=20pt plus 2pt minus 2pt
\baselineskip=16pt plus 2pt minus 2pt
%xxx
 {\count255=\time\divide\count255 by 60 \xdef\hourmin{\number\count255}
  \multiply\count255 by-60\advance\count255 by\time
  \xdef\hourmin{\hourmin:\ifnum\count255<10 0\fi\the\count255}}}
%       use \nolabels to get rid of eqn, ref, and fig labels in draft mode
\def\nolabels{\def\wrlabeL##1{}\def\eqlabeL##1{}\def\reflabeL##1{}}
\def\writelabels{\def\wrlabeL##1{\leavevmode\vadjust{\rlap{\smash%
{\line{{\escapechar=` \hfill\rlap{\sevenrm\hskip.03in\string##1}}}}}}}%
\def\eqlabeL##1{{\escapechar-1\rlap{\sevenrm\hskip.05in\string##1}}}%
\def\reflabeL##1{\noexpand\llap{\noexpand\sevenrm\string\string\string##1}}}
\nolabels
%
% tagged sec numbers
\global\newcount\secno \global\secno=0
\global\newcount\meqno \global\meqno=1
\def\newsec#1{\global\advance\secno by1\message{(\the\secno. #1)}
%\ifx\answ\bigans \vfill\eject \else \bigbreak\bigskip \fi  %if desired
\global\subsecno=0\eqnres@t\noindent{\bf\the\secno. #1}
\writetoca{{\secsym} {#1}}\par\nobreak\medskip\nobreak}
\def\eqnres@t{\xdef\secsym{\the\secno.}\global\meqno=1\bigbreak\bigskip}
\def\sequentialequations{\def\eqnres@t{\bigbreak}}\xdef\secsym{}
\global\newcount\subsecno \global\subsecno=0
\def\subsec#1{\global\advance\subsecno by1\message{(\secsym\the\subsecno. #1)}
\ifnum\lastpenalty>9000\else\bigbreak\fi
\noindent{\it\secsym\the\subsecno. #1}\writetoca{\string\quad 
{\secsym\the\subsecno.} {#1}}\par\nobreak\medskip\nobreak}
\def\appendix#1#2{\global\meqno=1\global\subsecno=0\xdef\secsym{\hbox{#1.}}
\bigbreak\bigskip\noindent{\bf Appendix #1. #2}\message{(#1. #2)}
\writetoca{Appendix {#1.} {#2}}\par\nobreak\medskip\nobreak}
%
%       \eqn\label{a+b=c}       gives displayed equation, numbered
%                               consecutively within sections.
%     \eqnn and \eqna define labels in advance (of eqalign?)
%
\def\eqnn#1{\xdef #1{(\secsym\the\meqno)}\writedef{#1\leftbracket#1}%
\global\advance\meqno by1\wrlabeL#1}
\def\eqna#1{\xdef #1##1{\hbox{$(\secsym\the\meqno##1)$}}
\writedef{#1\numbersign1\leftbracket#1{\numbersign1}}%
\global\advance\meqno by1\wrlabeL{#1$\{\}$}}
\def\eqn#1#2{\xdef #1{(\secsym\the\meqno)}\writedef{#1\leftbracket#1}%
\global\advance\meqno by1$$#2\eqno#1\eqlabeL#1$$}
%
%                        footnotes
\newskip\footskip\footskip14pt plus 1pt minus 1pt %sets footnote baselineskip
\def\footnotefont{\ninepoint}\def\f@t#1{\footnotefont #1\@foot}
\def\f@@t{\baselineskip\footskip\bgroup\footnotefont\aftergroup\@foot\let\next}
\setbox\strutbox=\hbox{\vrule height9.5pt depth4.5pt width0pt}
\global\newcount\ftno \global\ftno=0
\def\foot{\global\advance\ftno by1\footnote{$^{\the\ftno}$}}
%
%say \footend to put footnotes at end
%will cause problems if \ref used inside \foot, instead use \nref before
\newwrite\ftfile   
\def\footend{\def\foot{\global\advance\ftno by1\chardef\wfile=\ftfile
$^{\the\ftno}$\ifnum\ftno=1\immediate\openout\ftfile=foots.tmp\fi%
\immediate\write\ftfile{\noexpand\smallskip%
\noexpand\item{f\the\ftno:\ }\pctsign}\findarg}%
\def\footatend{\vfill\eject\immediate\closeout\ftfile{\parindent=20pt
\centerline{\bf Footnotes}\nobreak\bigskip\input foots.tmp }}}
\def\footatend{}
%
%     \ref\label{text}
% generates a number, assigns it to \label, generates an entry.
% To list the refs on a separate page,  \listrefs
%
\global\newcount\refno \global\refno=1
\newwrite\rfile
\def\ref{[\the\refno]\nref}
\def\nref#1{\xdef#1{[\the\refno]}\writedef{#1\leftbracket#1}%
\ifnum\refno=1\immediate\openout\rfile=refs.tmp\fi
\global\advance\refno by1\chardef\wfile=\rfile\immediate
\write\rfile{\noexpand\item{#1\ }\reflabeL{#1\hskip.31in}\pctsign}\findarg}
%       horrible hack to sidestep tex \write limitation
\def\findarg#1#{\begingroup\obeylines\newlinechar=`\^^M\pass@rg}
{\obeylines\gdef\pass@rg#1{\writ@line\relax #1^^M\hbox{}^^M}%
\gdef\writ@line#1^^M{\expandafter\toks0\expandafter{\striprel@x #1}%
\edef\next{\the\toks0}\ifx\next\em@rk\let\next=\endgroup\else\ifx\next\empty%
\else\immediate\write\wfile{\the\toks0}\fi\let\next=\writ@line\fi\next\relax}}
\def\striprel@x#1{} \def\em@rk{\hbox{}} 
\def\lref{\begingroup\obeylines\lr@f}
\def\lr@f#1#2{\gdef#1{\ref#1{#2}}\endgroup\unskip}

\def\addref#1{\immediate\write\rfile{\noexpand\item{}#1}} %now unnecessary
\def\startrefs#1{\immediate\openout\rfile=refs.tmp\refno=#1}
\def\xref{\expandafter\xr@f}\def\xr@f[#1]{#1}
\def\refs#1{\count255=1[\r@fs #1{\hbox{}}]}
\def\r@fs#1{\ifx\und@fined#1\message{reflabel \string#1 is undefined.}%
\nref#1{need to supply reference \string#1.}\fi%
\vphantom{\hphantom{#1}}\edef\next{#1}\ifx\next\em@rk\def\next{}%
\else\ifx\next#1\ifodd\count255\relax\xref#1\count255=0\fi%
\else#1\count255=1\fi\let\next=\r@fs\fi\next}
%

%
% this is ugly, but moore insists
\newwrite\ffile\global\newcount\figno \global\figno=1
\def\fig{fig.~\the\figno\nfig}
\def\nfig#1{\xdef#1{fig.~\the\figno}%
\writedef{#1\leftbracket fig.\noexpand~\the\figno}%
\ifnum\figno=1\immediate\openout\ffile=figs.tmp\fi\chardef\wfile=\ffile%
\immediate\write\ffile{\noexpand\medskip\noexpand\item{Fig.\ \the\figno. }
\reflabeL{#1\hskip.55in}\pctsign}\global\advance\figno by1\findarg}
\def\xfig{\expandafter\xf@g}\def\xf@g fig.\penalty\@M\ {}
\def\figs#1{figs.~\f@gs #1{\hbox{}}}
\def\f@gs#1{\edef\next{#1}\ifx\next\em@rk\def\next{}\else
\ifx\next#1\xfig #1\else#1\fi\let\next=\f@gs\fi\next}
\newwrite\lfile
{\escapechar-1\xdef\pctsign{\string\%}\xdef\leftbracket{\string\{}
\xdef\rightbracket{\string\}}\xdef\numbersign{\string\#}}

\def\writestop{\def\writestoppt{\immediate\write\lfile{\string\pageno%
\the\pageno\string\startrefs\leftbracket\the\refno\rightbracket%
\string\def\string\secsym\leftbracket\secsym\rightbracket%
\string\secno\the\secno\string\meqno\the\meqno}\immediate\closeout\lfile}}
\def\writestoppt{}\def\writedef#1{}
\def\seclab#1{\xdef #1{\the\secno}\writedef{#1\leftbracket#1}\wrlabeL{#1=#1}}
\def\subseclab#1{\xdef #1{\secsym\the\subsecno}%
\writedef{#1\leftbracket#1}\wrlabeL{#1=#1}}
\newwrite\tfile \def\writetoca#1{}
\def\leaderfill{\leaders\hbox to 1em{\hss.\hss}\hfill}
%       use this to write file with table of contents
\def\writetoc{\immediate\openout\tfile=toc.tmp 
   \def\writetoca##1{{\edef\next{\write\tfile{\noindent ##1 
   \string\leaderfill {\noexpand\number\pageno} \par}}\next}}}
%       and this lists table of contents on second pass

%
\catcode`\@=12 % at signs are no longer letters
%
%       Unpleasantness in calling in abstract and title fonts
\edef\tfontsize{\ifx\answ\bigans scaled\magstep3\else scaled\magstep4\fi}
\font\titlerm=cmr10 \tfontsize \font\titlerms=cmr7 \tfontsize
\font\titlermss=cmr5 \tfontsize \font\titlei=cmmi10 \tfontsize
\font\titleis=cmmi7 \tfontsize \font\titleiss=cmmi5 \tfontsize
\font\titlesy=cmsy10 \tfontsize \font\titlesys=cmsy7 \tfontsize
\font\titlesyss=cmsy5 \tfontsize \font\titleit=cmti10 \tfontsize
\skewchar\titlei='177 \skewchar\titleis='177 \skewchar\titleiss='177
\skewchar\titlesy='60 \skewchar\titlesys='60 \skewchar\titlesyss='60
\def\titlefont{\def\rm{\fam0\titlerm}% switch to title font
\textfont0=\titlerm \scriptfont0=\titlerms \scriptscriptfont0=\titlermss
\textfont1=\titlei \scriptfont1=\titleis \scriptscriptfont1=\titleiss
\textfont2=\titlesy \scriptfont2=\titlesys \scriptscriptfont2=\titlesyss
\textfont\itfam=\titleit \def\it{\fam\itfam\titleit}\rm}
 \ifx\answ\bigans\else scaled\magstep1\fi
\ifx\answ\bigans\def\abstractfont{\tenpoint}\else
\font\abssl=cmsl10 scaled \magstep1
\font\absrm=cmr10 scaled\magstep1 \font\absrms=cmr7 scaled\magstep1
\font\absrmss=cmr5 scaled\magstep1 \font\absi=cmmi10 scaled\magstep1
\font\absis=cmmi7 scaled\magstep1 \font\absiss=cmmi5 scaled\magstep1
\font\abssy=cmsy10 scaled\magstep1 \font\abssys=cmsy7 scaled\magstep1
\font\abssyss=cmsy5 scaled\magstep1 \font\absbf=cmbx10 scaled\magstep1
\skewchar\absi='177 \skewchar\absis='177 \skewchar\absiss='177
\skewchar\abssy='60 \skewchar\abssys='60 \skewchar\abssyss='60
\def\abstractfont{\def\rm{\fam0\absrm}% switch to abstract font
\textfont0=\absrm \scriptfont0=\absrms \scriptscriptfont0=\absrmss
\textfont1=\absi \scriptfont1=\absis \scriptscriptfont1=\absiss
\textfont2=\abssy \scriptfont2=\abssys \scriptscriptfont2=\abssyss
\textfont\itfam=\bigit \def\it{\fam\itfam\bigit}\def\footnotefont{\tenpoint}%
\textfont\slfam=\abssl \def\sl{\fam\slfam\abssl}%
\textfont\bffam=\absbf \def\bf{\fam\bffam\absbf}\rm}\fi
\def\tenpoint{\def\rm{\fam0\tenrm}% switch back to 10-point type
\textfont0=\tenrm \scriptfont0=\sevenrm \scriptscriptfont0=\fiverm
\textfont1=\teni  \scriptfont1=\seveni  \scriptscriptfont1=\fivei
\textfont2=\tensy \scriptfont2=\sevensy \scriptscriptfont2=\fivesy
\textfont\itfam=\tenit \def\it{\fam\itfam\tenit}\def\footnotefont{\ninepoint}%
\textfont\bffam=\tenbf \def\bf{\fam\bffam\tenbf}\def\sl{\fam\slfam\tensl}\rm}
\font\ninerm=cmr9 \font\sixrm=cmr6 \font\ninei=cmmi9 \font\sixi=cmmi6 
\font\ninesy=cmsy9 \font\sixsy=cmsy6 \font\ninebf=cmbx9 
\font\nineit=cmti9 \font\ninesl=cmsl9 \skewchar\ninei='177
\skewchar\sixi='177 \skewchar\ninesy='60 \skewchar\sixsy='60 
\def\ninepoint{\def\rm{\fam0\ninerm}% switch to footnote font
\textfont0=\ninerm \scriptfont0=\sixrm \scriptscriptfont0=\fiverm
\textfont1=\ninei \scriptfont1=\sixi \scriptscriptfont1=\fivei
\textfont2=\ninesy \scriptfont2=\sixsy \scriptscriptfont2=\fivesy
\textfont\itfam=\ninei \def\it{\fam\itfam\nineit}\def\sl{\fam\slfam\ninesl}%
\textfont\bffam=\ninebf \def\bf{\fam\bffam\ninebf}\rm} 
%
%---------------------------------------------------------------------
%
\def\noblackbox{\overfullrule=0pt}
\hyphenation{anom-aly anom-alies coun-ter-term coun-ter-terms}
\def\inv{^{\raise.15ex\hbox{${\scriptscriptstyle -}$}\kern-.05em 1}}

\def\Dsl{\,\raise.15ex\hbox{/}\mkern-13.5mu D} %this one can be subscripted
\def\dsl{\raise.15ex\hbox{/}\kern-.57em\partial}

\font\bigit=cmti10 scaled \magstep1
 %pound sterling
\def\lspace{\ifx\answ\bigans{}\else\qquad\fi}
\def\lbspace{\ifx\answ\bigans{}\else\hskip-.2in\fi} % $$\lbspace...$$
\def\boxeqn#1{\vcenter{\vbox{\hrule\hbox{\vrule\kern3pt\vbox{\kern3pt
        \hbox{${\displaystyle #1}$}\kern3pt}\kern3pt\vrule}\hrule}}}
\def\mbox#1#2{\vcenter{\hrule \hbox{\vrule height#2in
                \kern#1in \vrule} \hrule}}  %e.g. \mbox{.1}{.1}
%       matters of taste
%\def\tilde{\widetilde} \def\bar{\overline} \def\hat{\widehat}
%
% some sample definitions
  %     curly letters
    
   \def\CU{{\cal U}}
  \def\CD{{\cal D}}

\def\darr#1{\raise1.5ex\hbox{$\leftrightarrow$}\mkern-16.5mu #1}
 %pound sterling

 %puts a small half in a displayed eqn
\def\roughly#1{\raise.3ex\hbox{$#1$\kern-.75em\lower1ex\hbox{$\sim$}}}

\def\dwn{\Delta^*_{W_{n+1}}}

\def\nus{\nu^\star}

\def\hx#1{{\hat{#1}}}

\def\IW{{\bf WP}}
\def\Ds{\Delta^\star}
%%%
\def\abstract#1{
\vskip .5in\vfil\centerline
{\bf Abstract}\penalty1000
{{\smallskip\ifx\answ\bigans\leftskip 2pc \rightskip 2pc
\else\leftskip 5pc \rightskip 5pc\fi
\noindent\abstractfont \baselineskip=12pt
{#1} \smallskip}}
\penalty-1000}

\input epsf
%
%\draftmode
\noblackbox
\def\rla{\leftrightarrow}
\def\us#1{\underline{#1}}
\def\hth/#1#2#3#4#5#6#7{{\tt hep-th/#1#2#3#4#5#6#7}}
\def\nup#1({Nucl.\ Phys.\ $\us {B#1}$\ (}
\def\plt#1({Phys.\ Lett.\ $\us  {B#1}$\ (}
\def\cmp#1({Comm.\ Math.\ Phys.\ $\us  {#1}$\ (}
\def\prp#1({Phys.\ Rep.\ $\us  {#1}$\ (}
\def\prl#1({Phys.\ Rev.\ Lett.\ $\us  {#1}$\ (}
\def\prv#1({Phys.\ Rev.\ $\us  {#1}$\ (}
\def\mpl#1({Mod.\ Phys.\ Let.\ $\us  {A#1}$\ (}
\def\ijmp#1({Int.\ J.\ Mod.\ Phys.\ $\us{A#1}$\ (}
%%%
\def\br{\hfill\break}\def\ni{\noindent}
\def\cx#1{{\cal #1}}\def\al{\alpha}\def\IP{{\bf P}}
\def\tx#1{{\tilde{#1}}}\def\bx#1{{\bf #1}}
\def\ov#1#2{{#1 \over #2}}
\def\be{\beta}\def\al{\alpha}
\def\subsubsec#1{\ \br \noindent {\it #1} \br}

\def\zh{\hat{z}}
\def\lra{\leftrightarrow}

\def\dwn{\Delta^*_{W_{n+1}}}

\def\nus{\nu^\star}

\def\hx#1{{\hat{#1}}}

\def\IW{{\bf WP}}
\def\Ds{\Delta^\star}
%%%
\def\abstract#1{
\vskip .5in\vfil\centerline
{\bf Abstract}\penalty1000
{{\smallskip\ifx\answ\bigans\leftskip 2pc \rightskip 2pc
\else\leftskip 5pc \rightskip 5pc\fi
\noindent\abstractfont \baselineskip=12pt
{#1} \smallskip}}
\penalty-1000}
\def\us#1{\underline{#1}}
\def\hth/#1#2#3#4#5#6#7{{\tt hep-th/#1#2#3#4#5#6#7}}
\def\nup#1({Nucl.\ Phys.\ $\us {B#1}$\ (}
\def\plt#1({Phys.\ Lett.\ $\us  {B#1}$\ (}
\def\cmp#1({Comm.\ Math.\ Phys.\ $\us  {#1}$\ (}
\def\prp#1({Phys.\ Rep.\ $\us  {#1}$\ (}
\def\prl#1({Phys.\ Rev.\ Lett.\ $\us  {#1}$\ (}
\def\prv#1({Phys.\ Rev.\ $\us  {#1}$\ (}
\def\mpl#1({Mod.\ Phys.\ Let.\ $\us  {A#1}$\ (}
\def\ijmp#1({Int.\ J.\ Mod.\ Phys.\ $\us{A#1}$\ (}
%%%
\def\br{\hfill\break}\def\ni{\noindent}
\def\cx#1{{\cal #1}}\def\al{\alpha}\def\IP{{\bf P}}
\def\tx#1{{\tilde{#1}}}\def\bx#1{{\bf #1}}
\def\ov#1#2{{#1 \over #2}}
\def\be{\beta}\def\al{\alpha}
\def\subsubsec#1{\ \br \noindent {\it #1} \br}

\lref\FMW{R. Friedman, J.W. Morgan and E. Witten, \cmp 187 (1997) 679.}
\lref\MV{C. Vafa and D. Morrison, \nup 473 (1996) 74; \nup 476 (1996) 437.}
\lref\BM{P. Berglund and P. Mayr, {\it Heterotic string/F-theory duality
from mirror symmetry}, hep-th/9811217.}
\lref\FF{P. Mayr, \nup 494 (1997) 489.}
\lref\VW{C. Vafa and E. Witten, Nucl. Phys. Proc. Suppl. {$\us {46}$} 
(1996) 225.} 
\lref\KMV{S. Katz, P. Mayr and C. Vafa,
Adv. Theor. Math. Phys. $\us {1}$ (1998) 53.}
\lref\SDS{A. Klemm, 
W. Lerche, P. Mayr, C. Vafa, N. Warner,
                \nup 477 (1996) 746.}
\lref\KKV{S. Katz, A. Klemm and C. Vafa, \nup 497 (1997) 173.}
\lref\Loo{E. Looijenga, Invent. Math. {$\us {38}$} (1977) 17;
Invent. Math. {$\us {61}$} (1980) 1.}
%%%%%%%%%%%%%%%%%%%%%%%%%%%%%%%%%%%%%%%%%%%%%

\Title{\vbox{
\rightline{\vbox{\baselineskip12pt
\hbox{CERN-TH/98-39}
\hbox{hep-th/9904115}
}}}}
{$\cx N=1$ Heterotic String Vacua}
\vskip-1cm\centerline{{\titlefont from Mirror Symmetry} \foot{
Lectures presented at
the Winter School on Vector Bundles, Mirror Symmetry, 
and Lagrangian Submanifolds, Harvard University, March 1999}}\vskip 0.3cm
\centerline{%P. Berglund\foot{berglund@itp.ucsb.edu} and  
P. Mayr%\foot{Peter.Mayr@cern.ch}
}
\vskip 0.6cm
%\centerline{$^1$ \it Institute for Theoretical Physics, 
%University of California, Santa Barbara, CA 93106, USA}
%\vskip 0.0cm
\centerline{\it Theory Division, CERN, 1211 Geneva 23, 
Switzerland}
\vskip -0.8cm
\abstract{\ni
We review a systematic construction of 
$\cx N=1$ supersymmetric heterotic string vacua using 
mirror symmetry. The method provides a large class of 
explicit solutions for stable, holomorphic vector bundles
on Calabi--Yau $n$-folds $Z_n$ in terms of toric geometry.
Phenomenologically interesting compactifications as well
as non-perturbative dynamics of the heterotic string
are discussed within this framework.}
\Date{\vbox{\hbox{\sl {April 1999}}
}}\goodbreak

\parskip=4pt plus 15pt minus 1pt
\baselineskip=15pt plus 2pt minus 1pt

\newsec{Introduction}
Heterotic string compactifications with minimal $\cx N=1$ supersymmetry
have been the phenomenologically most promising candidate
for a unifying theory of particle physics and gravitation for
a while. In addition to having the smallest possible amount of 
supersymmetry, which is supposed to provide, amongst other, a
solution to the problem of why there is such a huge discrepancy between 
the scales determining the gravitational and gauge interactions, 
these theories have rich gauge symmetries, namely $E_8\times E_8$ or 
$SO(32)$, in ten dimensions. Those should be broken to 
the gauge groups of standard particle physics, be it grand
unified groups or that of the standard model.

The mathematical conditions that ensure $\cx N=1$ 
supersymmetry in a compactification on a compact manifold
times flat $d$-dimensional Minkowski space 
have been determined a long
time ago \ref\CHSW{P. Candelas, G.T. Horowitz, A. Strominger and
E. Witten, \nup 258 (1985) 46}\ref\HETS{
C. Hull and E. Witten, \plt 160 (1985) 398.;\br
E. Witten, \nup 268 (1986) 79.}\ref\DG{J. Distler,  \plt 188 (1987) 431;
J. Distler and B. Greene, \nup 304 (1988) 1.}. However explicit solutions to 
these equations are rather hard to obtain. In this lecture
we review a geometric construction of a large class of 
such heterotic vacua using a generalized version of mirror symmetry
in type II strings. In particular we define a map $f:\cx W_{n+1} \to (Z_n,V)$
that associates to a (toric) local $n+1$ dimensional Calabi--Yau geometry 
$\cx W_{n+1}$ the data of a (toric) Calabi--Yau $n$-fold $Z_n$ together with
a family of vector bundles $V$ (or more generally sheaves) on it.
The relation between the type II string compactifications
and  heterotic strings can be traced back to string dualities, namely
type IIA/heterotic \ref\HT{C.M. Hull and P.K. Townsend, \nup 438
(1995) 109}\ref\WDU{
E. Witten, \nup 443 (1995) 85}\ref\KV{
S. Kachru and C. Vafa, \nup 450 (1995) 69;\br
S. Ferrara, J.A. Harvey, A. Strominger and C. Vafa, \plt 361 (1995) 59.} 
and F-theory/heterotic 
duality \ref\vafaf{C. Vafa, \nup 469 (1996) 403}\MV. 
Within this context we establish this duality at the classic level and
provide a systematic construction of dual pairs $W_{n+1} \lra (Z_n,V)$,
where $W_{n+1}$ is now the global $n+1$-fold used for F-theory 
compactification.

\newsec{Heterotic $\cx N=1$ vacua}
The conditions for supersymmetric
compactifications of perturbative heterotic strings on a manifold
$Z$ have been determined 
in \CHSW\HETS\DG\
under some simplifying restrictions, such as the vanishing of 
the three-form  field strength $H=dB+\dots$ and a constant dilaton background.
These conditions are modified in non-perturbative vacua by the presence
of magnetic five-branes \ref\Ruben{M.J. Duff, R. Minasian and E. Witten,
\nup 465 (1996) 413.}.

\subsec{Perturbative vacua}
Supersymmetry in the uncompactified space-time requires the
existence of a covariantly constant spinor on $Z$ which means
that $Z$ is a Calabi--Yau manifold. In addition to the geometry
one has to specify a gauge background $V$ on $Z$ with the
following properties.

The vector bundle $V$ has to satisfy the topological conditions
\eqn\vtop{\eqalign{
c_1(V) &= 0\ {\rm mod}\ 2,\cr
\lambda(V) &= c_2(Z),}} 
where $\lambda$ is the four-dimensional characteristic class of
$V$.
The first equation ensures that the bundle $V$ admits spinors. 
The second equation derives from the Bianchi identity for the 
three-form field strength $H$.

Moreover supersymmetry requires that the connection on 
$V$ satisfies
\eqn\fsv{
F_{ab}=F_{\bar{a}\bar{b}}=0,\qquad g_{a\bar{b}}F^{a\bar{b}} = 0.}
So $V$ is holomorphic and moreover the
second equation says that $V$ is a stable bundle.  

The third Chern class of $V$ is also important: it determines
the net number of generations of the compactification as \DG
\eqn\matter{
N_{gen}-N_{anti-gen}={1 \over 2} c_3(V).}

Later it was realized that the relevant gauge backgrounds for string
theory are not actually smooth vector bundles but rather
include also more singular configurations such as reflexive or
coherent sheaves \ref\DGM{J. Distler, B.R. Greene 
and D. Morrison, \nup 481 (1996) 289.}. These are also the objects that 
appear naturally in the geometric construction discussed in the
following.

\subsec{Non-perturbative vacua: five-branes}
The understanding of non-perturbative dynamics of the heterotic
string improved dramatically with the uprise of string dualities.
The basic non-perturbative state of the heterotic theory
is the magnetic dual of the fundamental string, 
the five-brane. In particular, adding  background 
five-branes to a heterotic compactification can yield a new,
non-perturbative compactification with the same maximal amount
of supersymmetry as dictated by the compactification geometry \Ruben.
Non-perturbative five-branes add
an extremely rich spectrum of dynamics to the perturbative 
heterotic string, including non-critical strings 
\ref\ganor{O. Ganor and A. Hanany, \nup 474 (1996) 122;\br
N. Seiberg and E. Witten, \nup 471 (1996) 121.} and 
non-abelian gauge symmetries of very 
high rank \ref\BIAMC{J.D. Blum
and K. Intriligator, \nup 506 (1997) 199; \nup 506 (1997) 223;\br
P.S. Aspinwall and D.R. Morrison,
\nup 503 (1997) 533;\br
P. Candelas, E. Perevalov and G. Rajesh \nup 507 (1997) 445.}.

The presence of non-perturbative five-branes changes the 
topological constraint on the gauge background $V$.
The magnetic five-brane contributes a source term 
to the Bianchi identity for the three-form $H$:
\eqn\bibi{
dH = tr R\wedge R - tr F\wedge F- q_{5B}\sum_{five-branes} \delta^{(4)}_{5B},}
which integrates on a four-cycle to 
\eqn\bibii{
c_2(Z)=c_2(V)+[W].}
Here $\delta^{(4)}_{5B}$ is a formal four-form that integrates 
to one in the directions {\it transverse} to a single five-brane
and $[W]$ denotes the class associated to this last term in \bibi.
In other words, a five-brane is able to correct the mismatch of gravitational
and gauge curvatures in the direction transverse to the world volume.
Thus adding five-branes, which take care of part of the 
anomaly related to $c_2(Z)$, enlarges substantially the
class of consistent gauge backgrounds on $Z$ \FMW. This 
interplay between gauge fields and five-branes is clear from
the interpretation of the latter as the zero size limit 
of gauge instantons \ref\witsmi{E. Witten, \nup 460 (1996) 541.}\ganor.

Lorentz invariant compactifications with five-branes are
possible in $d\leq 6$ compactifications. In six dimensions,
the curvature is localized along K3 and
the five-branes simply fill space-time. 
These five-branes contribute also to the massless spectrum.
On the five-brane of the $E_8 \times E_8$ string, there is 
an antisymmetric tensor $B^\prime$ with anti-self-dual field strength.
It couples to anti-self-dual strings in 6d with a string tension
parametrized by a real scalar $\phi^\prime$. There are four 
more real scalars $\phi_i$ which parametrize the position of the 
five-brane in the transverse space. There is also an $SO(5)_R$ R-symmetry
under which $\phi^\prime,\phi_i$ transform as a $\us 5$ and
$B^\prime$ as a singlet.

In four uncompactified dimensions the five-brane world volume
has to wrap a supersymmetric 2-cycle $C_2$ in $Z$, that is 
a holomorphic curve \ref\SSC{K. Becker, M. Becker and 
A. Strominger, \nup 456 (1995) 130}\ref\vafaii{M. 
Bershadsky, V. Sadov and C. Vafa, \nup 463 (1996) 420.}.
The massless spectrum depends now on the geometry of the 2-cycle 
$C_2$. The answer can be obtained from the twisting of the 
world volume theory \vafaii. It would be interesting to 
obtain the general answer. Here we
restrict to some simple interesting situations.
A special supersymmetric twisting for a five-brane compactified
on a curve $C_2^{(g)}$ of genus $g$ has been considered in \SDS\ with
the result that out of the five scalars, two become 1-forms
on $C_2^{(g)}$ while three remain scalars $\tilde{\phi_i}$. In this case 
we get $g$ vector fields from $B^\prime$ with one index in the
$C_2^{(g)}$ direction, as well as $2g+3$ real scalars and a single
anti-symmetric tensor $B^\prime$ in four dimensions\foot{Note that it was 
the non-compactness of $C_2$ in \SDS\ that kills the scalars from 0-forms
on $C_2^{(g)}$.}.

If $g=1$, the normal bundle of $C_2^{(1)}$ is trivial and the
above analysis applies. There are six scalars and a single vector field
which combine to a $\cx N=4$ vector multiplet.
This situation is encountered if $C_2^{(1)}$ is the fiber of the elliptic
fibration of $Z$.
Enhancement to $SU(N)$ takes place if $N$ five-branes coincide.

On the other hand consider the situation where $C_2$ is an 
isolated genus 0 curve in $Z$. Since $B^\prime$ is a 
singlet, it is unaffected by the twisting and must survive,
similarly $\phi^\prime$ which still parametrizes the
tension of the four-dimensional string to which $B^\prime$ couples.
Since $C_2$ is isolated there are no further scalars. The
spectrum is that of an $\cx N=1$ linear multiplet. 

Apart from the generic spectrum for generic positions of 
the five-branes there will be additional light degrees of
freedom for special values of the moduli. If several
five-branes come close to each other,
non-critical string with a vanishing tension will become relevant,
similarly as in six dimensions.
In the case of $g>1$, winding states of the string 
contribute massless vector multiplets, charged with respect
to the generic gauge fields from $B^\prime$ as in \SDS. 

Another feature is the presence of extra matter in the
perturbative heterotic gauge group for a special alignment
of the gauge and geometric moduli \BM. This matter of
non-perturbative origin is in addition to the contribution
in \matter.

\subsubsec{Elliptically fibered Calabi--Yau manifolds $Z$}
Generally, finding a suitable gauge background $V$ fulfilling
\vtop\ and \fsv\ explicitly is a very hard question, though
there is an existence theorem for a solution in the generic case 
\ref\UY{K. Uhlenbeck and S.T. Yau, {\it On the existence of 
hermitian Yang--Mills connections in stable vector bundles},
preprint (1996).}.
There is a much better understanding in the case where $Z$ is elliptically
fibered, basically because it is relatively simple to describe
holomorphic (semi-)stable bundles on a torus
$E$ and fibering these data in a holomorphic
way over a complex manifold $B$ produces a holomorphic bundle $V$
on the elliptic fibration $E\to Z \to B$. This situation has been
considered in \FMW\ref\BJPS{M. Bershadsky, 
A. Johansen, T. Pantev and V. Sadov,
\nup 505 (1997) 165.}%
\ref\DO{R.Y. Donagi, Asian J. Math. $\us 1$ (1997) 214;
MSRI pub. 28 (1992) 65.}. 

In the following we will take a quite
different route to the subject of holomorphic stable vector 
bundles (or rather sheafs)
on such $n$-dimensional Calabi--Yau manifolds $Z_n$. We
define them purely
in terms of type IIA string theory.
The fact that the non-abelian 
gauge symmetries of type II arise from D-brane geometries will
lead us to a unified (toric) geometric description of bundles on $Z_n$ 
in terms of type IIA compactification geometries. 

\newsec{Vector bundles and type II strings}
Instead of the heterotic string, we will consider now the ten-dimensional 
type IIA string compactified on Calabi--Yau $n+1$-fold singularities. Though
it is not obvious at first sight, this theory has the
same moduli space $\cx M_V$ of holomorphic stable bundles on 
a (different) Calabi--Yau manifold $Z_n$.

We will derive this statement in two steps, for $n=2$ and $n>2$.
For $n=2$ we have a relation between Wilson lines on a $T^2$ and
Calabi--Yau 2-fold singularities which follows from a symmetry 
of the type IIA compactification on K3 $\times T^2$ that is
closely related to mirror symmetry of K3. 
In the second step 
we use the adiabatic principle \VW\ and consider holomorphic
fibrations of the 2-fold singularities to construct
Calabi--Yau $n+1$-fold singularities related to 
holomorphic stable bundles on Calabi--Yau $Z_n$.

\subsec{Mirror symmetry on K3 and generalizations}
\subsubsec{Mirror symmetry of K3}
Consider a type IIA string compactified on a K3 surface. 
This gives an $\cx N=2 $ supersymmetric theory in six dimensions.
From dimensional reduction  we obtain twenty 
$\cx N=2$ vector multiplets which contain  as their bosonic degree 
of freedom each one vector and four real scalars. The latter
transform as a $\us{4}$ under an $SO(4)_R$ R-symmetry and parametrize the 58 metric moduli of K3
plus 22 values of the $B$-fields on $H^2(K3)$.  

In a a given algebraic realization $M_2$ of K3, 
the 80 moduli of K3 split
in complexified K\"ahler and complex structure moduli\foot{Subscripts,
as in $M_2$,
denote the complex dimension of a geometry.}. The
$SO(4)_R$ rotations acting on the scalars in $\us{4}$ do not
preserve this split in general. In particular there is a $SO(4)_R$
rotation that exchanges K\"ahler and complex structure moduli spaces
$\cx M_{KM}$ and $\cx M_{CS}$, respectively. Such a transformation
is usually known as {\it mirror symmetry} of K3 manifolds, which 
says that a type IIA string compactified on $M_2$ describes the same
physics as a type IIA string compactification on a "different"\foot{Since
the K3 surface is unique, mirror symmetry is actually a discrete 
identification in the moduli space of K3 surfaces.} K3
manifold $W_2$.
The new algebraic 
K3 surface $W_2$ is called the mirror manifold of $M_2$. We have the
following identifications under mirror symmetry:
\eqn\msi{\eqalign{
M_2 &\rla W_2,\cr
\cx M_{CS}(M_2) &\rla \cx M_{KM}(W_2),\cr
\cx M_{KM}(M_2) &\rla \cx M_{CS}(W_2).}}

\subsubsec{An extension of mirror symmetry}
\def\cxms{\cx M^{string}_E}
There is an intriguing generalization of this symmetry upon 
compactification on a further $T^2$ to four dimensions. 
In this case we obtain a $\cx N=4$ supersymmetric string theory.
The $\cx N=4$ vector multiplet contains now 6 scalars 
which transform in a $\us 6$ of an $SO(6)_R$ symmetry.
The two extra scalars are the internal components $A_i$ of the six-dimensional
gauge fields on the $T^2$, call it $E$,
and thus parametrize the stringy moduli space of
Wilson lines on the elliptic curve, $\cxms(H)$. Here $H$ denotes
the structure group of the Wilson line background. Again the 
$SO(6)_R$ rotations provide identifications within the type IIA
moduli space $\cx M_{IIA}(K3\times T^2)$:
\eqn\msii{
\cx M_{KM}(M_2) \lra \cx M_{CS}(W_2) \lra \cxms(H).}
In particular there is now an element of $SO(6)_R$, corresponding
to the last arrow, which provides a
relation between
the moduli space of stringy
Wilson lines with the K\"ahler moduli space of an algebraic
K3 manifold. 

\subsubsec{The dual heterotic picture}
Let us illustrate the symmetry \msii\ of the type IIA compactification on 
K3 $\times T^2$ in the dual heterotic picture where it is very obvious;
however we stress that we do not need the heterotic dual to derive the
symmetry. Fig. 1 shows the type IIA compactification
and the corresponding dual heterotic string compactified on 
$T_1^2\times T_2^2\times T_3^2$. The factorization of $T^6$ is related to
$i)$ the elliptic fibration and $ii)$ the factorization of the gauge 
background on K3 $\times T^2$ in the type IIA theory. Under these circumstances, the K\"ahler moduli of K3 can be identified with moduli space $\cx M_{T^2_1}(H_1)$
of $H_1$ Wilson lines on the first $T^2_1$ and 
similarly complex structure moduli
with Wilson lines in the second factor $\cx M_{T^2_2}(H_2)$.
The non-abelian gauge symmetry in six dimensions is $G$ which on the
type IIA side arises from the $G$ singularity and on the heterotic side is
the commutant of $H_1\times H_2$ in $G^{het}_0$. Upon further compactification
to four dimensions on a torus $T^2_3$, we obtain another factor 
$\cx M_{T^2_3}(G)$ on both sides.
\vskip 0.5cm
{\baselineskip=12pt \sl
\goodbreak\midinsert
\centerline{\epsfxsize 3.5truein\epsfbox{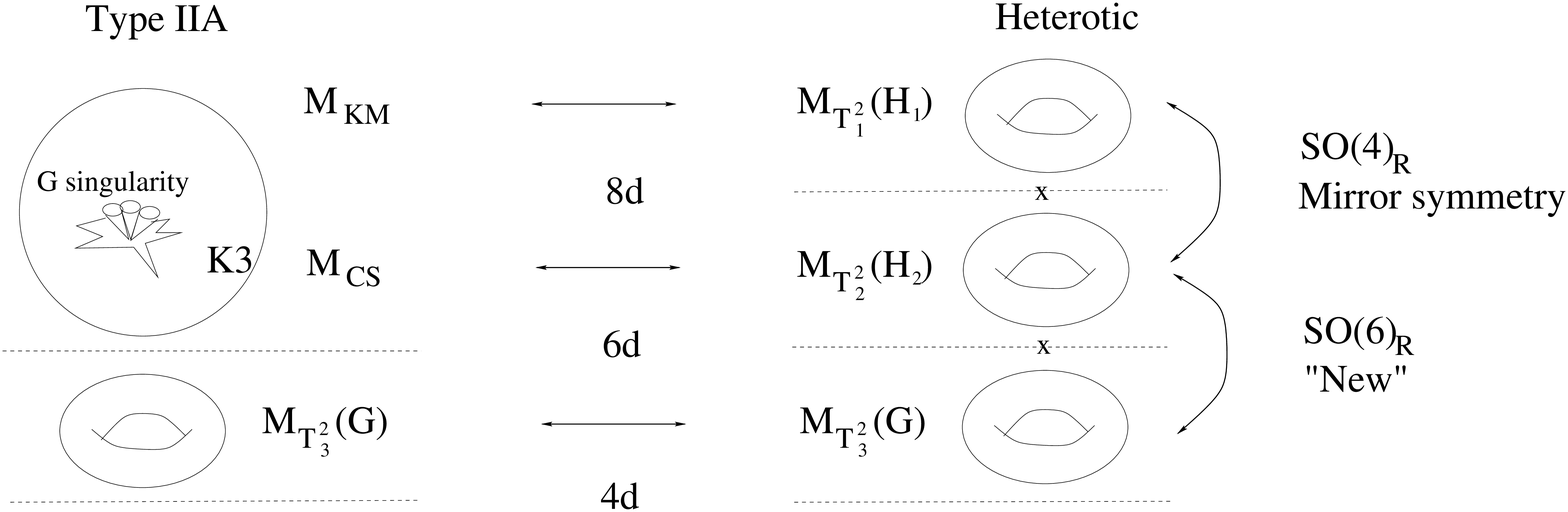}}
\leftskip 1pc\rightskip 1pc \vskip0.3cm
\noindent{\ninepoint  \baselineskip=8pt 
{{\bf Fig. 1:}
The symmetry \msii\ from a heterotic point of view.}
}\endinsert}\vskip -0.4cm
\ni
The heterotic theory has an obvious symmetry under permutation of the 
three $T^2$ factors. The exchange of the first two tori amounts to
mirror symmetry in the type IIA theory. On the other hand an exchange
that involves the third torus gives a new symmetry of the moduli
space of the type IIA compactification that identifies Wilson lines
with structure group $G$ on $T^2$ with either K\"ahler or complex
deformations of the K3. This is the new relation described in \msii.

\subsec{The precise correspondence}

We want now to use the equivalence \msii\ implied by the $SO(6)_R$
transformations to describe the moduli space $\cx M_E(H)$ of flat holomorphic 
(and non-abelian) $H$ bundles on an 
elliptic curve in terms of complex deformations
of a geometric type IIA compactification.
There are a few basic considerations to make precise such a correspondence. 
We will not dive into the details here, which can be found
in \KMV\BM, but rather sketch the procedure and state the result.

\item{a)}
Firstly note that the field theoretical moduli space $\cx M_E(H)$
contains also the deformations of the torus $E$. This is related to
the fact that we have to consider gauge backgrounds for the $K3\times T^2$
compactification which factorize in the sense that they do not restrict
the geometry of the $T^2$ factor. 

\item{b)} The moduli space $\cxms(H)$
that appears in \msii\ is still that of a theory of Wilson lines coupled to
{\it string} theory. To obtain the field theoretical moduli space $\cx M_E$
we have to decouple the string states by sending the string scale 
to infinity, $M_{str}
\to \infty$. 

\item{c)} The structure group $H$ of the Wilson lines on $T^2$ is
contained in the gauge group of the six-dimensional compactification 
on $M_2$, which is generically abelian. Non-abelian 
gauge symmetries in six dimensions arise from charged vector multiplets
associated to D2 brane wrappings on 2-cycles $C^i$ in $M_2$ \WDU.  
The masses of these states are $\sim M_{str} \cdot Vol(C^i)$.
Therefore light non-abelian vector multiplets are associated to very
small 2-cycles in $M_2$ in the limit $M_{str}\to \infty$.

The condition $a)$ translates to the property that $M_2$ has to be 
elliptically fibered. Conditions $b)$ and $c)$ are well-known from 
the geometric engineering of $\cx N=2$ quantum field theories 
\ref\GE{A. Klemm, 
W. Lerche, P. Mayr, C. Vafa, N. Warner,
                \nup 477 (1996) 746;\br
S. Katz, A. Klemm and C. Vafa,
\nup 497 (1997) 173;\br
S. Katz, P. Mayr and C. Vafa,
Adv. Theor. Math. Phys. $\us {1}$ (1998) 53.}.
Taking the string decoupling limit and keeping at the same time
the masses of the non-abelian states finite results in a geometric
limit of K3 which is captured by a {\it local singularity} of
the K3. Singularities of K3 at finite distance in the moduli
space are the well-known  ADE singularities. Their local 
versions are described in terms of ADE singularities of ALE spaces. 

The local geometry is then of the form of an elliptic fibration over the
$\bx C$ plane with an $H$ singularity, with $H\subset ADE$, at the origin.
There are $r={\rm rank}\ H$ 2-spheres $C^i$ associated to the blow up 
of this singularity which intersect according to an ordinary $H$
Dynkin diagram. The elliptic fibration implies the existence of a
further 2-sphere $C^0$ associated to the generic homology class of
the elliptic fiber; the intersections with the $r$ 2-cycles $C^i$
are described by the {\it affine} Dynkin diagram of $H$. 
In summary we arrive at the following equivalence \KMV\BM:

\item{o} Let $\cx M_2$ denote the local neighbourhood of an $H$
singularity in an  elliptic fibration over the plane. There are $r+1$
volumes associated to $C^k,\ k=0,\dots,r$ which parametrize
the K\"ahler moduli space $\cx M_{KM}(\cx M_2)$ of $\cx M_2$. 
The type IIA compactification on $\cx M_2$ describes holomorphic
stable $H$ bundles on $E$ with a moduli space 
parametrized by the $r+1$ K\"ahler moduli of $\cx M_2$: 
\eqn\equi{M_{KM}(\cx M_2) \cong  M_{E}(H).}

\ni
Applying local mirror symmetry to the geometry $\cx M_2$ we 
obtain an equivalent description in terms of complex structure deformations
of the mirror geometry $\cx W_2$:

\item{o} Let $\cx W_2$ denote the (local) mirror of $\cx M_2$.
The type IIA compactification on $\cx W_2$ describes holomorphic
stable $H$ bundles on $E$ parametrized by the $r+1$ complex moduli
of $\cx M_2$:
\eqn\equi{M_{CS}(\cx W_2) \cong  M_{E}(H).}
\goodbreak

\subsec{The complex geometries $\cx W_2$}
Let us briefly summarize the structure of the two-complex
dimensional geometries $\cx W_2$ which are the mirrors of
the local ADE singularities in the elliptic fibration 
over the plane. We will recover in this way the results of
\FMW\ for those $H$ where a geometric realization
of $\cx M_E(H)$ has been given and reinterpret them 
in terms of type IIA compactifications. Moreover
we obtain a unified and general geometric formulation for any $H$.

The geometry $\cx W_2$ is given as a hypersurface defined by
the vanishing of a quasi-homogeneous polynomial in some
variables $(y,x,z,v)$ on a (toric) ambient space. Its general
form is 
\eqn\gfi{p_{\cx W_2}=p_0(y,x,z)+p_+(y,x,z;v)=p_0+\sum_{i=1} v^ip^i_+(y,x,z)=0,}
where we have separated the $v$ independent term $p_0$,
and all polynomials are quasi-homogeneous in $(y,x,z)$.
The $v$ independent term $p_0$ is universal for all $H$ and given by
\eqn\pn{p_0=y^2+x^3+z^6+\mu yxz.}
$p_0=0$ describes an elliptic curve with complex structure modulus $\mu$.
The $v$ dependent term depends on the structure group 
$H$ and has an intriguing group
theoretical structure:  consider the affine Dynkin diagram $\Gamma(H)$ 
associated to the Lie group $H$. Then each node in $\Gamma(H)$ with Dynkin
index $s_i$ contributes one monomial to $p^i_+(y,x,z)$. See 
appendix A for an illustration of this fact.

Let us give one simple example. For $H=A_{N-1}$, $\Gamma(A_{N-1})$
contains $N$ nodes with Dynkin index one. The equation defining
$\cx W_2(A_{N-1})$ obtained from local mirror symmetry (for $N$ even) is
\eqn\gfii{
p_0+v\; (a_Nz^N+a_{N-2}z^{N-2}x+a_{N-3}z^{N-3}y+\dots+a_0x^{N/2})=0,}
with $p_0$ as above. 
This defines a two complex dimensional local geometry with a
complex structure determined by the complex parameters $\mu,a_k$.
In fact $\mu$ defines the complex structure of the torus $\hat{E}:p_0=0$
on which the bundle is defined. The ${a_k}$ are coordinates on
the $\IP^{N-1}$ predicted in the mathematics literature 
\Loo.
It is well-known, that as far as complex structure is concerned,
linear variables in the defining "superpotential" can be simply integrated
out, which amounts to setting $dW/dv=0$ in addition to $W=0$. 
In our case, varying with respect to $v$ gives 
$p_+=0$ and $p_0=0$ separately.
This is the zero-dimensional spectral cover description found in \FMW.

Moreover, from the above construction, we get a similar description
for all structure groups $H$ in terms of complex deformations of
a two-dimensional complex geometry $\cx W_2(H)$ which is identified
as a physical type IIA compactification geometry. 

Interestingly, we can even describe more sophisticated bundles,
such as the $Spin(32)/\bx Z_2$ bundles without vector structure
considered in 
\ref\Witwov{E. Witten, J. High Energy Phys. $\us{02}$ (1998) 6.},
in terms of two-complex dimensional geometries $\cx W_2$.
These bundles lead to a disconnected set of vacua for 
the heterotic $Spin(32)/\bx Z_2$ string, the CHL vacua \ref\chl{S. 
Chaudhuri, G. Hockney and J. D. Lykken,
\prl 75 (1995) 2264.} with maximal supersymmetry but reduced rank of the
gauge group. The bundles associated to these vacua can be constructed
in terms of special elliptic fibrations with reduced monodromy
group of the fibration \ref\BKMT{P. Berglund, A. Klemm, P. Mayr 
and S. Theisen, 
{\it On type IIB vacua with varying coupling constant}, hep-th/9805189.}.

\subsec{Vector bundles on Calabi--Yau $n$-folds}
Apart from the satisfying formulation of holomorphic $H$ bundles
on a torus, unifying and extending the results obtained in \FMW,
the real pay-off of the geometric description derives from the
type IIA compactification picture. Combining the above
construction with an adiabatic argument as in \ref\VWi{C. Vafa and 
E. Witten, Nucl. Phys. Proc. Suppl. {$\us {46}$} 
(1996) 225.}, we can 
get local type IIA compactifications that describe holomorphic
stable vector bundles on Calabi--Yau $n$-folds in a simple way.
All we have to do is to fiber the two-dimensional geometries
$\cx W_2$ holomorphically over a $n-1$ dimensional base $B_{n-1}$,
such that the total geometry describes a local patch of a 
type IIA compactification on a Calabi--Yau $n+1$-fold $\cx W_{n+1}$.
The general form of $\cx W_{n+1}$ is as in \gfi, with the
difference that the complex parameters defining the
complex structure, i.e. $\mu,a_k$ in the $A_{N-1}$ case,
now get upgraded to sections of line bundles on the base $B_{n-1}$.
In particular the $v$ independent term $p_0$ now describes an
elliptically fibered Calabi--Yau $n$-fold $Z_n$, rather than
a torus, and $p_+$ defines a family of holomorphic stable
bundles on $Z_n$.
The fiber construction can be done most easily within the framework
of toric geometry as will be described in sect.6.

\newsec{Heterotic/type IIA and heterotic/F-theory duality}
We started the discussion of stable holomorphic vector bundles
on Calabi--Yau $n$-folds in the framework of heterotic 
$\cx N=1$ string vacua, and arrived at the same physics 
in the last section using only type IIA physics. In particular
we are free to interpret these type IIA data as valid classical heterotic
vacua. We thus arrive at the conclusion that the two theories
are the same in the proposed limits - the local mirror limit of type IIA
and the perturbative low energy limit of heterotic string.
This is the "classical" part of a conjectured duality 
of full quantum string theories, namely
between type IIA on a Calabi--Yau $n+1$-fold $W_{n+1}$ and 
heterotic string compactified on a $Z_n\times T^2$, with 
$Z_n$ a  Calabi--Yau $n$-fold. Note that we used 
only type IIA physics to derive our results and therefore 
proved such an equivalence on the classical level 
rather than assuming or using it.

Actually we would like to get rid off the $T^2$ factor on the heterotic
side. The fact that we used elliptically fibered geometries\foot{This
requires that the mirror geometries $\cx W_2$ of the elliptically
fibered geometries $\cx M_2$ are also elliptically fibered. This
is the case.} makes it
possible to take what is called the F-theory \vafaf\ limit: the limit in 
which the K\"ahler class of the elliptic fiber shrinks to zero size\foot{See 
also \ref\Asi{P. Aspinwall, {\it K3 surfaces and string duality},
hep-th/9611137}.}.
Note that this limit can be taken in $\cx W_{n+1}$ without interfering
with our construction since we concentrated on the complex deformations, 
rather than the K\"ahler deformations.

Note that the geometric construction of holomorphic stable bundles
on Calabi--Yau $n$-folds $Z_n$ in terms of local geometries
works for any rank of the structure group $H$. 
However precisely if $H$ fits into 
the primordial gauge symmetry of the heterotic string, that is
$E_8\times E_8$ or $SO(32)$, the local geometries $\cx W_{n+1}$
can be embedded in global compact Calabi--Yau's $W_{n+1}$.
In this case, from the point of F-theory/heterotic 
string duality, we have obtained a systematic way to construct dual pairs: 
we associate to the Calabi--Yau $n+1$ fold $W_{n+1}$
used for F-theory compactification the heterotic data of a Calabi--Yau $n$-fold
$Z_n$ and a bundle on it (and {\it vice versa}).

\newsec{Holomorphic stable bundles on elliptic Calabi--Yau manifolds}
In the next section we will sketch the precise formulation of the
geometric construction of holomorphic stable bundles on 
elliptic Calabi--Yau manifolds in terms of toric geometry. 
Before doing that, we outline some qualitative properties of these bundles
and the geometric approach to them.

\subsec{Some known results}
We proceed with an informal collection of some known facts about holomorphic
stable vector bundles on Calabi--Yau manifolds $Z$. The interested
reader can find the details in the literature, as 
quoted. Most of the concepts in the following sections can be 
followed without being familiar with the definitions and facts 
presented in this paragraph.

\def\itemb{\item{$\bullet$}}
\itemb A holomorphic stable bundle $V$ on an elliptic curve $E$ is
simply a flat bundle on $E$. It is specified by a representation
of the abelian fundamental group of $E$ in a maximal torus of 
the structure group $H$ (assuming the latter is simply connected). 

\itemb \Loo:
The moduli space  of flat $H$ bundles on a fixed
$E$ is a weighted projective space $\IW^r_{s_0,\dots,s_r}\equiv W^r$.
Here $r={\rm rank}\ H$ and $s_i$ are the Dynkin indices of the 
affine Dynkin diagram associated to $H$ (see appendix A.). 

\itemb {\FMW: A geometric realization of the moduli space 
$\cx M_E$ of flat bundles on $E$ (including the deformations of $E$)
can be given as follows:
for $H=A_{N-1}$, $V$ is specified by $N$ unordered
points on the Jacobian $\hat{E}$   
of $E$ (corresponding to the values of the Wilson lines in the fundamental
representation). If $E$ is given in Weierstrass form
\eqn\torus{
p_0=y^2+4x^3-g_2x-g_3=0,}
the $N$ points can be defined as the roots of a degree $N$ polynomial
\eqn\sc{
p_C=a_N+a_{N-2}x+a_{N-3}y+\dots+a_0x^{N/2}=0}
for $N$ even, the last term being $a_0yx^{(N-3)/2}$ for $N$ odd.
The coefficients $a_k$ parametrize the $\IP^{N-1}$ predicted in \Loo.\br
\indent For $H=E_k,\ k=6,7,8,$ a geometric description can be
obtained from  del Pezzo surfaces $dP_k$. E.g., for $k=8$,
the equation 
\eqn\dP{\eqalign{y^2=&
4x^3-g_2xv^4-g_3v^6+(\al_6u^6+\al_5u^5v+\al_4u^4v^2
+\al^3u^3v^3+\al^2u^2v^4)\cr&+x(\be_4u^4+\be_3u^3v+\be_2u^2v^2+\be_1uv^3),}}
describes a del Pezzo surface obtained from blowing up eight
points in $\IP^2$. The elliptic curve is identified with the 
hyperplane $u=0$. The coefficients $\al_k,\be_k$ 
parametrize the $\IW^{8}_{1,2,2,3,3,4,4,5,6}$ 
predicted in \Loo. 
}

\itemb \FMW: Holomorphic stable bundles on elliptically fibered Calabi--Yau
manifold $Z$ can be described by fibering the data $(E,W^r)$ holomorphically
over a complex base $B$. The non-trivial fibration of the torus $E$ over $B$
builds up the Calabi--Yau $n$-fold $Z_n$. Similarly,
the projective spaces $W^r$ fit into a 
holomorphic bundle $\us W^r$ over $B$. If $s:B\to \us W^r$ is
a section, the homogeneous coordinates $\tilde{a}_i$ of $\us W^r$
pull back to sections
\eqn\omb{
\tilde{a}_i\in H^0(B,\cx N^{s_i}\otimes \cx L^{-di}),}
where $\cx L$ is the anti-canonical bundle of $B$ and
$\cx N$ a line bundle on $B$ which is an important characteristic of 
$V$ and is closely related to the higher Chern classes of $V$. 
The $d_j$ are the degrees of the independent Casimir operators of 
the group $H$ and $s_i$ the Dynkin indices as above. 
Part of the data of $V$ are defined by choosing a section of $\us W^r$.
\br
\indent
The space $\cx Y$ of sections of $\us W^r$ is the base of the
moduli space of holomorphic flat bundles over $Z$; the fiber data
include additional continous and discrete data related to certain
twistings of $V$\foot{See
\ref\CD{G. Curio and R.Y. Donagi, \nup 518 (1998) 603.} for a partial
identification of these data and refs. \FMW\BJPS
\ref\CU{G. Curio, \plt 435 (1998) 39.} for comments on the
data related to additional singularities in the construction.}.
 \br
\indent 
A universal abstract construction of $V$, the so-called 
parabolic construction, has been given in terms of deformations 
of unstable bundles. The method is so far restricted to 
bundles that are invariant under the involution $y\to -y$ on the
fiber and have $c_3(V)=0$.

\itemb \FMW: Chern classes of $V$ have been calulated in the
parabolic and spectral cover approach for $H=SU(N)$: 
\eqn\dodi{\eqalign{
c_2(V)&=\eta\sigma - {1 \over 24} c_1(B)^2(N^3-N)
-{1\over 8} N \eta (\eta -Nc_1(B)) -{1\over 2} \pi_*(\gamma^2).}}
Here $\pi$ denotes the elliptic fibration $\pi:Z\to B$, $\sigma$
the class of the section, $\eta=c_1(\cx N)$
and $\gamma=\lambda(N\sigma-\eta+Nc_1(B))$ is a certain 
class in $H^{1,1}(C,\bx Z)$ which fulfills $\pi_* \gamma=0$. 
Here $C$ is the spectral cover 
described by an equation of the form \sc\ with $a_i$ replaced by 
$\tilde{a}_i$ as in \omb\ and moreover $\lambda \in \bx Z/2$.

The third Chern class $c_3(V)$ has been determined in 
\ref\BA{B. Andreas, {\it On vector bundles and chiral matter 
in N=1 heterotic compactifications}, hep-th/9802202}\CU:
\eqn\ciii{
c_3(V) = 2\lambda\sigma\eta(\eta-Nc_1(B)).}

\subsec{The geometric approach}
The geometric type II string approach improves in various aspects 
the previous understanding and use of holomorphic stable 
bundles and heterotic/F-theory duality. We will obtain

\itemb
A canonical description of flat bundles $V$ on a torus $E$ in terms of 
two-complex dimensional type IIA compactification geometries,
provided by a map $f:\ \cx W_2 \to (E,V)$.
This unifies the results of \FMW\ for $A_N$ and $E_k$ and
generalizes to any structure group $H$.

\itemb 
More generally we obtain 
a map $f:\ \cx W_{n+1} \to (Z_n,V)$ that assigns to a local Calabi--Yau
$n+1$-fold $\cx W_{n+1}$ an elliptically fibered 
Calabi--Yau $n$-fold $Z_n$ together with
a family of holomorphic stable vector bundles $V$ on it.

\itemb
All objects will be defined within the framework of toric geometry.
Differently than in the methods summarized above, 
singularities in the elliptic fibration 
of $Z_n$ and sheaf generalizations of $V$ are automatically
taken care of in the toric formulation. 

\itemb
On the practical side, the method allows us to freely engineer 
bundles (sheafs) on any elliptically fibered Calabi--Yau 
manifold $Z_n$.
The defining data, such as the structure group $H$, the manifold
$Z_n$ and characteristic classes of $V$ 
have an extremely simple representation in terms of the toric
polyhedron associated to the geometry $\cx W_n$. 

\itemb 
Interpreting the data $(Z_n,V)$ in terms of the heterotic string,
one obtains valid $\cx N=1$ vacua that solve \fsv\bibi. 
The coincidence of the type IIA physics on $\cx W_n$ and 
heterotic string on $(Z_n,V)$ is related to the conjectured
non-perturbative type IIA/heterotic and F-theory/heterotic 
string dualities. We use classical type IIA
physics to derive this equivalence, rather than assuming it.
This elucidates type IIA/heterotic duality and 
F-theory/heterotic duality at the classical level.

\itemb
Extending the map $f$ in a global context 
$f:\ W_{n+1} \lra (Z_n,V)$ provides a systematic identification 
of dual pairs of F-theory/heterotic duality.

\itemb
Important data of the bundle $V$, such as the line bundle $\cx N$,
stability of $V$ as well as non-perturbative five-branes, have a very 
transparent interpretation in the toric formulation.

\newsec{Toric geometry: A do-it-yourself kit for
(non-perturbative) heterotic $\cx N=1$ vacua}
After outlining the conceptual framework in the previous sections,
we want now to describe a concrete recipe, how to define and construct
the complex manifold $W_{n+1}$ such that it describes a chosen 
"heterotic" Calabi--Yau $Z_n$ and a gauge background $V$ on it\foot{Since
we consider structure groups $H$ which are embeddable in the heterotic
gauge group we will treat the local version
$\cx W_{n+1}$ and the global embedding $W_{n+1}$ on the
same footing in the following.}.
We will define the dual pair  $W_{n+1}$ and $(Z_n,V)$ within
the framework of toric geometry, which for many purposes is the
most useful and general description of Calabi--Yau manifolds.
We will see that toric geometry is also quite suited to 
describe vector bundles on Calabi--Yau's \foot{A fact that has been
noticed already in the formulation of $(0,2)$ linear sigma 
models \DGM.}.

The toric construction of the local geometries is described in
\BM\ and we will not repeat the details here. However the
toric description has some very simple and graphical aspects
which we want to sketch here. Actually we can formulate a 
set of quite simple rules using the language of toric polyhedra.
In particular we will see how the physical quantities,
such as the structure group $H$, the "heterotic" manifold $Z_n$,
data of the characteristic classes of $V$ appearing in \FMW,
singularities of the gauge background as well as associated
non-perturbative five-brane wrappings are encoded in a
"toric polyhedron".

In a nutshell, toric geometry describes a $k$ 
dimensional Calabi--Yau geometry $M_k$ by
a $k+1$ dimensional polyhedron $\Ds_{k+1}$, defined as 
the convex hull of a set $\{\nus_i\}$ of $r$ integral vertices $\nus_i$ 
defined in a standard
$k+1$ dimensional lattice. Each vertex $\nus_i$ on a face of $\Ds_{k+1}$
of codimension larger than one defines a 
divisor (holomorphic hypersurface) in $M_k$. Note that because of
the dimension of $\Ds_{k+1}$, there must be $r-k-1$ 
linear relations $l^{(m)}_i$
between the $r$ (external) vertices $\nus_i$
\eqn\mori{
\sum_i l^{(m)}_i\nus_i=0,\ m=1,\dots,h^{1,1}.}
In fact $h^{1,1}=r-k-1$ is the number of K\"ahler moduli of $M_k$.

Now choose the elliptically fibered "heterotic" Calabi--Yau $n$-fold $Z_n$ 
with base $\pi:Z_{n}\to B_{n-1}$ of the fibration. To $Z_n$ 
corresponds an $n+1$ dimensional polyhedron $\Ds_{Z_n}$. The 
guideline to construct families of holomorphic stable bundles
on $Z_n$ in terms of toric geometry is then as follows \BM:

\item{1.}{\bf The structure group $H$}:\br
Choose the structure group $H$. The polyhedron $\Ds_3$ for the
geometry $\cx W_2(H)$ and, if existing, a global embedding $W_2(H)$ is given
in \KMV\BM. The perturbative gauge symmetry of the heterotic string
is the commutant of $H$ in $E_8\times E_8$.

\item{2.} {\bf The base geometry}: \br
The toric base of the elliptic fibration $B_{n-1}$
is described by a toric polyhedron $\Ds_{B_{n-1}}$ as given 
e.g. in \BM.

\item{3.} {\bf The fibration $W_2 \to W_{n+1} \to B_{n-1}$}:\br
The two polyhedra are now embedded in a $3+n-1$ 
dimensional polyhedron $\Ds_{n+2}$ that describes a Calabi--Yau 
$W_{n+1}$:
\eqn\polyi{\Ds_{W_{n+1}} \subset \pmatrix{0 &\Ds_3 \cr \Ds_{B_{n-1}}&*}.}

\item{4.} {{\bf Embedding of $Z_n$}:\br
To  the heterotic compactification manifold 
$Z_n$ corresponds a $n+1$ dimensional polyhedron 
$\Ds_{Z_n}$. It appears
as a {\it projection} in $\Ds_{W_{n+1}}$. To define the projection we note
that the fact that we use elliptically 
fibered geometries implies \ref\AKMS{A. C. Avram, M. Kreuzer,
M. Mandelberg and H. Skarke, \nup 494 (1997) 567.} that
$\Ds_3$ is itself of a specific form. In our case $\Ds_3$ splits as 
\eqn\polyii{\Ds_3=\pmatrix{0&\Ds_2\cr\pm 1& *},}
where $\Ds_2$ is the two-dimensional polyhedron defining the elliptic
fiber. The projection to the heterotic manifold is defined 
in the direction transverse to the hyperplane of the elliptic fiber:
\vskip 0.5cm
{\baselineskip=12pt \sl
\goodbreak\midinsert
\centerline{\epsfxsize 0.9truein\epsfbox{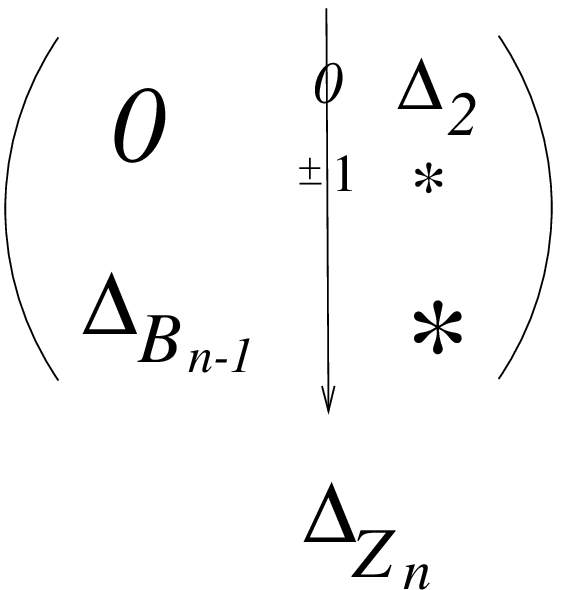}}
\leftskip 1pc\rightskip 1pc \vskip0.3cm
\noindent{\ninepoint  \baselineskip=8pt 
%{{\bf Fig. 1:}
%Dynkin diagrams for the duals of the untwisted Kac-Moody algebras. 
%The integers denote the associated Dynkin labels for the affine root.}
}\endinsert}\vskip -0.4cm
This concludes the description of the geometrical data of the heterotic
compactification\foot{
Note that there can be obstructions to give all curves
in $Z_n$ a finite size; this phenomenon is related to non-perturbative
gauge symmetries \BM.}.}

\item{5.}{{\bf  Perturbative bundle data: 
fixing the line bundle $\cx N$ on $ B_{n-1}$}:\br
The above conditions do not fix $\Ds_{W_n+1}$
completely: there are many $n+2$ dimensional vertices that 
project to the $n+1$ dimensional vertices of $\Ds_{Z_n}$ in
the above construction. This freedom is related to perturbative
and non-perturbative data associated to the gauge background $V$.
Let us first consider the perturbative data. According to \FMW,
the definition of $V$ includes the specification of a line bundle 
$\cx N$ on $B_{n-1}$. The class $\eta=c_1(\cx N)$ enters the 
Chern classes $c_2(V)$ and $c_3(V)$ as in \dodi\ciii.
\br
\indent In the toric framework, line bundles are related to the linear 
relations \mori\ between the 
vertices of $\dwn$. Recall also that $\dwn$ is required
to be convex. Making a minimal choice of vertices $\{\nus_i\}$,
such that $\Ds_{W_{n+1}}$ fulfils the conditions { 1.-4.},
and moreover is convex, determines obviously the linear relations 
\mori\ of the vertices in 
$\Ds_{W_{n+1}}$ and thus in particular the line bundles on
$B_{n-1}$. Specifically it turns out that
there are only two line bundles that determine
the fibration of the two complex-dimensional geometries $W_2$ over
$B_{n-1}$, namely the 
anti-canonical bundle $\cx L$ of $B_{n-1}$ and one additional
line bundle which is equivalent to  $\cx N$ of ref.\FMW.\br 
\indent 
Thus determining the bundle data $\cx N$, translates in the toric 
polyhedra to a simple linear relation of the vertices in $\dwn$.
In fact the definition in toric geometry is more general in the sense
that it applies to any singularities in the geometry or the gauge
background. An interesting fact is that the convexity of $\Ds_{W_{n+1}}$
restricts the possible linear relations and thus the range of acceptable
line bundles $\cx N$. In six dimensions this phenomenon describes 
the well-known stability properties of instantons on K3 and we 
expect a similar interpretation in four dimensions. A general bound
on the first Chern class $\eta=c_1(\cx N)$ is given by
\eqn\ebound{
\nu(G)\  c_1(\cx L) \leq \eta \ \ (\leq 12 c_1(\cx L)\ ),
} where $G$ is the singularity type of the fiber geometry $W_2$ and
$\nu(G)$ is a certain characteristic number defined in \ref\BMii{P. Berglund
and P. Mayr, {\it Stability of vector bundles from F-theory}, hep-th 9904114.}
Explicit expressions $\cx N$ for toric bases, such as $\IP^2$ and $\bx F_n$
and blow ups thereof, can also be found in \BM.
See also 
ref.\ref\RA{G. Rajesh, J. High Energy Phys. $\us{12}$ (1998) 18.} for an independent discussion of the class $\eta$
in terms of toric polyhedra\foot{The bound on $\eta$ derived in \RA\ is
different from \ebound.}.}

\item{6.} {{\bf  Non-perturbative bundle data I: 
gauge symmetries} :\br
After fixing $H$, $Z_n$ and the line bundle $\cx N$, there is
still a freedom to add vertices to $\dwn$. This freedom
corresponds to introducing singularities in the behavior
of $V$ that are related to non-perturbative physics of the
heterotic string. Specifically we can introduce
singularities in the elliptic fibration of $W_{n+1}$ located over a
divisor of $B_{n-1}$. Gauge symmetry enhancement follows from 
the physics of the type IIA/F-theory compactification on $W_{n+1}$ \MV.
In the heterotic picture these gauge symmetries are non-perturbative.
Amazingly, the simple requirement of convexity of $\dwn$ gives 
an almost complete information about the possible complicated non-perturbative 
gauge symmetries, e.g. in the case of five-branes at singularities \BIAMC.
We will describe some singularities of the heterotic bundle associated
to this non-perturbative gauge enhancements in the next section.}

\item{7.} {{\bf  Non-perturbative bundle data II: non-perturbative five-branes} :\br
We can also blow up the base $\tx B_{n}$ of the {\it elliptic} 
fibration of $W_{n+1}$. It is a $\IP^1$ bundle over the heterotic 
base $B_{n-1}$. In the six dimensional case, corresponding
to  $n=1$ and a K3 compactification of the heterotic string,
blow ups of the $\IP^1$ fibration have been identified as 
non-perturbative heterotic 5 branes filling space time and 
located at points of the K3 \MV. The story in four dimensions is similar \BM,
with the difference that the five-branes are  now wrapped on holomorphic
spheres $C$ in $B_{n-1}$ specified by the position of the blow-up\foot{See
also ref. \RA\ for a discussion of these five-branes.}.
For $C$ a genus zero curve the universal spectrum from the 
five-branes is an $\cx N=1$ tensor multiplet $L$. 
It couples to
a non-critical string in four dimensions. This is a quite interesting
object from the point of supersymmetric $\cx N=1$ particle 
physics. In particular the string tension can be classically
zero but acquires a small, non-perturbatively generated 
tension $\sim \Lambda^{-2}$ at the quantum
level \FF, very similarly as the non-perturbative mass of the
$W^\pm$ bosons in the $\cx N=2$ $SU(2)$ Seiberg--Witten model.
It would be interesting to study their implication for standard
particle physics below the scale $\Lambda$.}

\ni
This concludes our recipe for the construction of the
heterotic $\cx N=1$ vacuum. The precise information about which 
toric vertices implement the points 1.-7. can be found
in \BM. It is striking how simple toric concepts, such as 
convexity of $\dwn$, contain interesting physics. Moreover
singularities, both of the geometry $Z_n$ and  the bundle $V$
can be easily introduced in the toric language.
However
we should also mention that there are few additional data of
$V$ \FMW\ which are not encoded in the simple toric description 
above. It would be interesting to have a generalization that
contains them.

Note that apart from the classical geometric formulation of 
(families of) holomorphic stable gauge backgrounds on $Z_n$,
we can think of the correspondence $W_{n+1}\lra (Z_n,V)$ 
as a dual pair and assume, use - or possibly verify - 
duality of the {\it quantum} 
theories, to study quantum and non-perturbative phenomena (as
we did in fact already,  when we used F-theory to infer non-perturbative
heterotic phenomena). Some interesting directions that have been
considered include correlation functions and the $\cx N=1$ 
superpotential. See refs. 
\ref\QSD{E. Witten, \nup 474 (1996) 343;
P. Mayr, \nup 494 (1997) 489;
S. Katz and C. Vafa, \nup 497 (1997) 204;
C. Vafa, Adv.Theor.Math.Phys. $\us 2$ (1998) 497;
W. Lerche, J. High Energy Phys. $\us{11}$ (1997) 4;
W. Lerche and S. Stieberger, 
{\it Prepotential, mirror map and F theory on K3}, hep-th/ 9804176;
W. Lerche, S. Stieberger and N.P. Warner, hep-th/9811228; hep-th/9901162.}
for examples.

\newsec{Some results}
Let us finally sketch some results on (non-perturbative) heterotic
physics which have been 
obtained in \BM\ using the above method. 

\subsec{Standard embedding}
A simple solution to the equations \vtop,\fsv\ is to set the
gauge connection equal to the spin connection of the manifold,
$V=TZ$. Though this configuration is not too appealing for
phenomenology - the heterotic gauge group is $E_6\times E_8$ and
the matter spectrum is tied to the Euler number of $Z$ - it 
has been studied extensively in the past because it was one
of the few known solutions. The structure group of $V$ is then 
$SU(n)$ for a Calabi--Yau $n$-fold, and the local geometry 
described by an equation of the form \gfii.

The global geometry $W_{n+1}$ corresponding to the F-theory dual 
can be determined in the following way. From the fact
that the bundle in the second factor is trivial it follows that 
the line bundle $\cx N$ discussed in sect.6 is trivial 
in the second $E_8$ factor. The Weierstrass form for $W_{n+1}$ and $Z_n$ 
takes the following form:
\eqn\tbiii{\eqalign{
p_{Z_n}&=y^2+x^3+x\tx z^4f+\tx z^6g,\cr
p_{W_n}&=y^2+x^3+x(\zh z w)^4f+(\zh z w)^6g+\zh^6z^5w^7\Delta+\zh^6z^7w^5.}}
Here $(z,w)$ denote the variables parametrizing the base $\IP^1$
of the elliptically fibered K3 fiber $W_2$ of the 
fibration $\pi_F:W_{n+1}\to B_{n-1}$ and 
$\Delta=4f^3+27g^2$
is the discriminant of the elliptic fibration of $Z_n$.
This generalizes the six-dimensional result 
obtained in \ref\AD{P.S. Aspinwall and R.Y. Donagi, {\it
The Heterotic string, the tangent bundle, and derived
                  categories}, hep-th/9806094.} to four and lower dimensions.

\subsec{Non-perturbative gauge symmetries}
In the F-theory picture gauge symmetries
correspond to singularities in the elliptic fibration. Those located
over the base $B_{n-1}$ have a perturbative interpretation in the
heterotic dual, whereas non-perturbative ones are located over 
curves in $B_{n-1}$ \MV. Using the map
$W_{n+1}\to (Z_n,V)$ we can associate the heterotic compactification 
that creates the same dynamics non-perturbatively . In the six-dimensional 
case we find:

\leftskip .5cm \rightskip .5cm
\ni {\it  $(*)$
Consider the $E_8\times E_8$ string compactified on 
an elliptically fibered K3  with a singularity of type 
$G$ at a point $s=0$ and a special gauge background $\hx V$. If
the restriction of the spectral cover\foot{For $G\neq SU(N)$ we use
the generalizations of the spectral cover discussed 
previously.} 
of $\hx V_{|E}$ to the fiber $E$ at $s=0$ is
sufficiently trivial, the heterotic string acquires a 
non-perturbative gauge symmetry $G_{np} \supset G$. }
\vskip0.2cm

\leftskip .0cm \rightskip .0cm
\ni
The triviality 
condition can be made precise by specifying
the behavior of $V$ near $s=0$ \BM. Similar results hold
for four-dimensional compactifications.

We point out that the above triviality condition on the spectral cover
does not imply that the field strength of $V$, which measures the behavior
of $V$ near the singularity, is trivial.
In fact it has been shown recently in 
\ref\witkt{E. Witten, 
{\it Heterotic string conformal field theory and A-D-E singularities},
hep-th/9909229.}
that if $F=0$
on the singularity, the conformal field theory of the heterotic
string is well behaved and there is no non-perturbative gauge
symmetry.

\subsec{Non-perturbative dualities}
Our map $f:\ W_{n+1} \to (Z_n,V)$ can be ambiguous in the sense
that there are two (or even more) ways to associated a pair
$(Z_n,V)$ to $\dwn$. If the two maps are compatible with
the same elliptic fibration, we obtain a non-perturbative duality
of two heterotic string theories
\eqn\duaii{
(Z_n,V)\sim (Z_n^\prime,V^\prime).}
The conditions under which
this duality exists, can be formulated in simple properties of
the polyhedron $\dwn$ associated to $W_{n+1}$ (essentially
the existence of appropriate hyperplanes and projections in $\dwn$) \BM.
The generic form of the duality is the following:

\leftskip 0.5cm \rightskip 0.5cm\ni
{\it $(**)$ Let the heterotic string be compactified  
on a Calabi--Yau three-fold with $G^\prime$ singularity and 
with a certain gauge background with structure group  $H$
such that the toric data $\dwn$ fulfil the above mentioned condition.
Then there exists a 
non-perturbatively equivalent compactification
on a Calabi--Yau manifold 
with $G$ singularity and with a specific gauge background
with structure group  $H^\prime$}. 
\vskip0.1cm

\leftskip 0.cm \rightskip 0.cm\ni
Here $H$ $(H^\prime)$
is the commutant of $G$ $(G^\prime)$ in $E_8\times E_8$. Note that
the duality exchanges the groups of the geometric singularity and the
gauge bundle: the singularity in the dual manifold is the commutant of
the structure group and {\it vice versa}.
Let
us give an extreme example of such an duality: 
the heterotic string with a trivial gauge bundle on a smooth
K3 has a perturbative $E_8\times E_8$ gauge symmetry and
in addition $n_T^\prime=24$ non-perturbative tensor multiplets
from the 24 five-branes required to satisfy \bibii. The dual theory
is a heterotic theory with $E_8\times E_8$ gauge bundle on 
a K3 with $E_8\times E_8$ singularity. The perturbative
gauge symmetry is trivial, while non-perturbative dynamics
associated to the singularity produce both, the $E_8\times E_8$
gauge symmetry as well as 24 extra tensor multiplets.

This duality reminds very much of a known duality in the linear sigma model
formulation of heterotic strings,
namely a symmetry of the formulation under the exchange
of the data defining the manifold and the data defining the bundle
\ref\DK{J. Distler and S. Kachru, \nup 442 (1995) 64.}.  
However note that in our case this duality is 
in general {\it non-perturbative}.
 
\subsec{Mirror symmetry of F-theory}
Consider a six-dimensional F-theory compactification on $W_3$
and the associated heterotic dual $(Z_2,V)$ as defined above.
If the mirror manifold $M_3$ of $W_3$ is also elliptically
fibered, there is also a heterotic theory $(Z_2^\prime,V^\prime)$
associated to it and one can ask the question of how the two theories 
are related. The answer is that after compactification on a 
three-torus to three dimensions, the two become equivalent in virtue
of mirror symmetry of type II strings 
\ref\IS{K. Intriligator and N. Seiberg, \plt 387 (1996) 513.}.

A proposal for the relation between $(Z_2,V)$ and $(Z_2^\prime,V^\prime)$
has been made in \ref\PR{E. Perevalov and G. Rajesh, \prl 79 (1997) 2931.}
based on a comparison of hodge numbers 
and gauge symmetries in some cases. Using the toric map
$W_{n+1}\to (Z_n,V)$ one can derive the two theories; in fact one finds
a subtle realization of Higgs and Coulomb branches in toric geometry
as expected from the action of three-dimensional mirror symmetry \BM.
The general relation is of a similar form as in $(**)$ above:
a heterotic theory with a $G$ bundle compactified on a K3 manifold
with $\hx H$ singularity gets mapped to a heterotic theory 
with a bundle with structure group $\hx H$ compactified on a
manifold with $G$ singularity. Note that we have an explicit
map of the moduli spaces of the two theories. It would be interesting
to formulate the associated three-dimensional dual pairs
associated to this relation, as in 
\ref\HOV{K. Hori, H. Ooguri and C. Vafa, \nup 504 (1997) 147.}.

\newsec{Outlook}
In these lectures we have reviewed a construction of a certain class of 
Calabi--Yau geometries $\cx W_{n+1}$ and, if possible, 
their compact embeddings 
$W_{n+1}$. In a certain limit of the type
IIA string compactified on $W_{n+1}$ we obtain 
the physics of holomorphic, stable vector bundles (or sheaves) $V$ 
on a Calabi--Yau
manifold $Z_n$. Interpreting the data $(Z_n,V)$ as a valid classic
heterotic vacuum establishes F-theory/heterotic duality at a classical 
level. In the toric construction, we have a well-defined toric map 
$f:\ W_{n+1}\to (V,Z_n)$ that allows for a systematic construction of 
dual pairs in both directions.

The identifications between $W_{n+1}$ and $(Z_n,V)$ may serve as a 
starting point to study more refined non-perturbative relations 
between the dual F- and heterotic string theory. In particular we
would like to formulate both, perturbative and non-perturbative 
quantities of the heterotic string, e.g. gauge and Yukawa couplings of 
the $(0,2)$ vacua, in terms of geometric quantities on $W_{n+1}$,
such as the correlation functions determined by period integrals 
considered in \FF. In the ideal (and supposedly too optimistic) case, 
we can hope to obtain exact results for at least some of the holomorphic 
physical quantities in the $\cx N=1$ theory from geometric information
on $W_{n+1}$, similarly as it happened to work in the 
case of $\cx N=2$ supersymmetry
\KV\ref\PPL{
S. Kachru, A. Klemm, W. Lerche, P. Mayr and C. Vafa, \nup 459 (1996) 537.}\GE.

$$ \ $$

\noindent{\bf Acknowledgements}: \br
The author would like to thank the organizers for the opportunity
to present this lectures and their hospitality and P. Berglund, S. Katz and
C. Vafa for the pleasant collaboration on the subjects presented above.

\vfill
\goodbreak

\vbox{\ 
\vskip -1.4cm
\centerline{\bf Appendix A. Local mirror geometries ${\cal W}_2$}
On the right, the affine Dynkin diagrams together with the Dynkin indices
$s_i$ are shown. The polynomial $p_{\cx W_2}=0$ for the 
associated local complex geometry $\cx W_2$ is shown on the left. 
Each node in the Dynkin diagram of index $s_i$ contributes precisely
one monomial with $v$ power $s_i$ to $p_{\cx W_2}$.

\vskip 0.4cm

\def\v{{\bf v}}
\def\ov#1#2{#1 \over #2}
\def\ep{\epsilon}
\def\mquad{\qquad\qquad\qquad\qquad\qquad\qquad
\qquad\qquad\qquad\qquad\qquad\qquad}

{\goodbreak\midinsert
\rightline{\epsfxsize 1.6truein\epsfbox{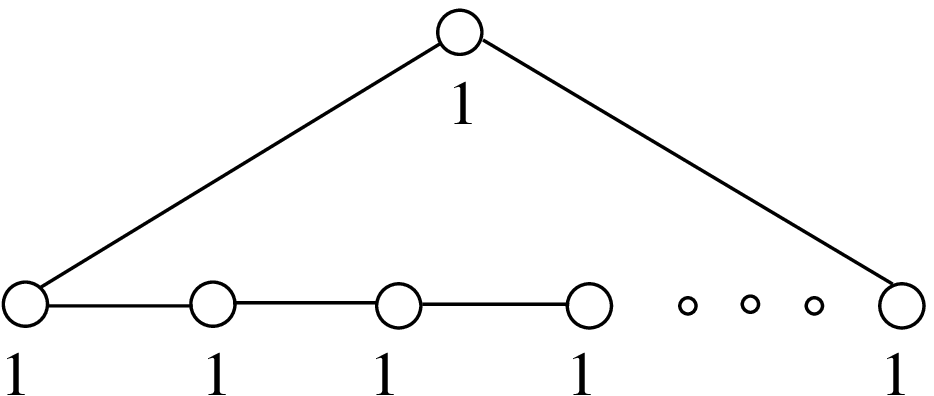}}
\endinsert}
\vskip -2.7cm\leftskip -2cm
$$\eqalign{
{\bf A_{N-1}}:\ \ & \v^0\ (y^2+x^3+z^6+  yxz)\ +\mquad \cr
& \v^1\ (z^N +z^{N-2}x +z^{N-3}y +\dots+{x^{N/2}\brace yx^{\ov{N-3}{2}}})
}$$

\vskip 0.5cm
{\goodbreak\midinsert
\rightline{\epsfxsize 2truein\epsfbox{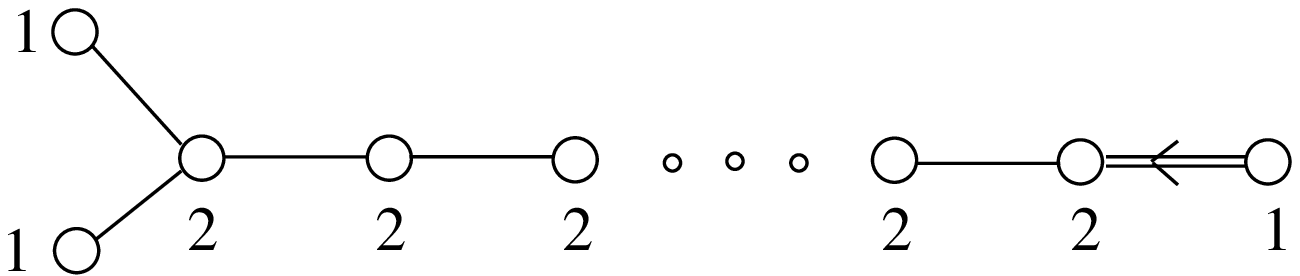}}
\endinsert}
\vskip -2.4cm\leftskip -2cm
$$\eqalign{{\bf B_N:}\ \ & \v^0\ (y^2+x^3+z^6+  xyz)\ + \mquad\cr
&\v\ (z^N+z^{N-3}y+\ep yx^{(N+1)/2-2}
+(1-\ep)x^{N/2})\ + \cr
 &\v^2\ (z^{2N-6}+z^{2N-8}x+z^{2N-10}x^2+\dots+x^{N-3})
}
$$

\vskip 0.5cm
{\goodbreak\midinsert
\rightline{\epsfxsize 1.6truein\epsfbox{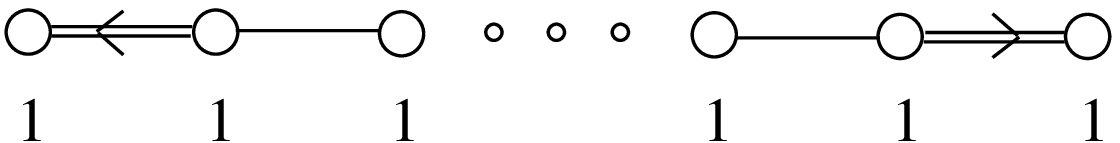}}
\endinsert}
\vskip -2cm\leftskip -2cm
$$\eqalign{
{\bf C_{N}:}\ \ & \v^0\ (y^2+x^3+z^6+  yxz)\ +\mquad\cr
&\v\ (z^{2N} +z^{2N-2}x +z^{2N-4}x^2 +\dots+x^{N})}
$$

\vskip 0.5cm
{\goodbreak\midinsert
\rightline{\epsfxsize 2.2truein\epsfbox{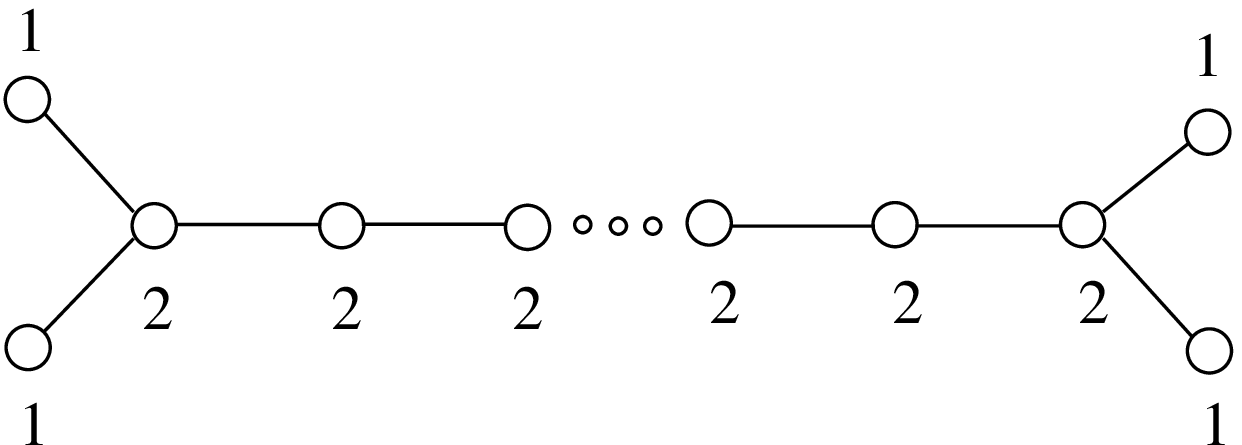}\hskip -0.8cm}
\endinsert}
\vskip -3.3cm\leftskip -2cm
$$\eqalign{{\bf D_N:}\ \ &\v^0\  (y^2+x^3+z^6+  xyz)\ +\mquad \cr
&\v\ (z^{N-3}y+z^{N}+z^{1-\ep}yx^{(N+\ep)/2-2}
+z^{\ep}x^{(N-\ep)/2})\ + \cr
 &\v^2\ (z^{2N-6}+z^{2N-8}x+z^{2N-10}x^2+\dots+z^2x^{N-4})
}
$$

\vskip 0.6cm
{\goodbreak\midinsert
\rightline{\epsfxsize 0.75 truein\epsfbox{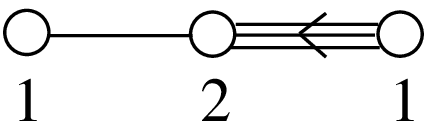}\hskip 1cm}
\endinsert}
\vskip -2cm\leftskip -2cm
$$\eqalign{
{\bf G_{2}:}\ \ & \v^0\ (y^2+x^3+z^6+  yxz)+\mquad \cr
&\v\ (z^3+y)\ +\ \v^2}
$$

\vskip 0.6cm
{\goodbreak\midinsert
\rightline{\epsfxsize 1.4 truein\epsfbox{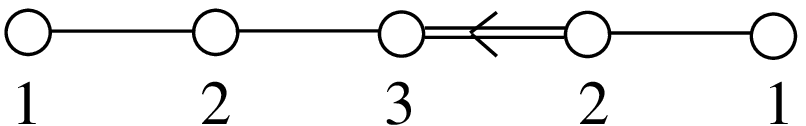}\hskip 0.3cm}
\endinsert}
\vskip -2cm\leftskip -2cm
$$\eqalign{{\bf F_{4}:}\ \ 
& \v^0\ (y^2+x^3+z^6+  yxz)\ +\mquad \cr
&\v\ (z^4+x^2)\ +\ \v^2\ (z^2+x)\ +\ \v^3}
$$

\vskip 0.5cm
{\goodbreak\midinsert
\rightline{\epsfxsize 1.1 truein\epsfbox{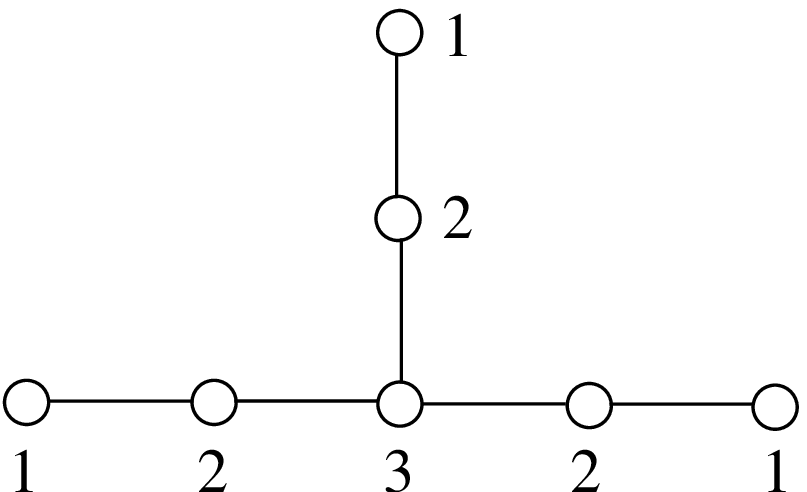}\hskip 0.6cm}
\endinsert}
\vskip -2.5cm\leftskip -2cm
$$\eqalign{{\bf E_6}:\ \ 
&\v^0\ (y^2+x^3+z^6+xyz)\ +\mquad\cr
&\v\ (z^5+zx^2+xy)\ +\ \v^2\ (z^4+z^2x+zy)\ +\ \v^3\ z^3
}$$

\vskip 0.6cm
{\goodbreak\midinsert
\rightline{\epsfxsize 1.8 truein\epsfbox{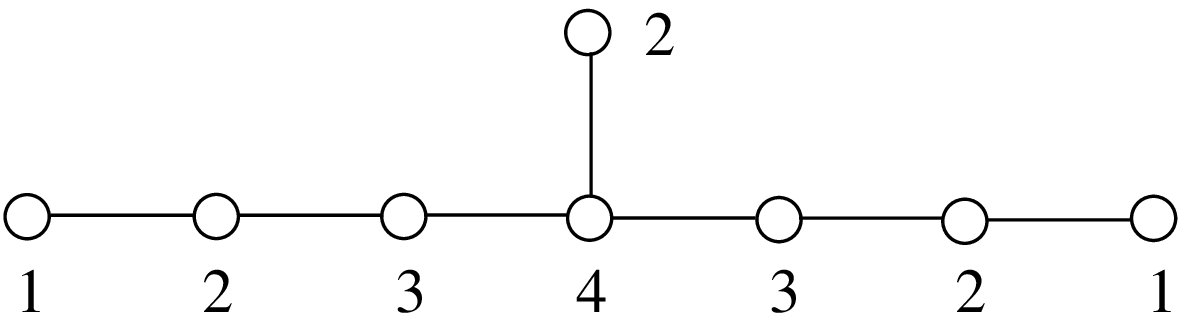}\hskip -0.5cm}
\endinsert}
\vskip -2.5cm\leftskip -2cm
$$\eqalign{{\bf E_7}:\ \ 
& \v^0\ (y^2+x^3+z^6+xyz)\ +\mquad \cr 
&\v\ (z^5+xy)\ +\ \v^2\ (z^4+zy+x^2)\ +\ \v^3\ (z^3+zx)\ +\ \v^4\ z^2
}$$

\vskip 0.6cm
{\goodbreak\midinsert
\rightline{\epsfxsize 2 truein\epsfbox{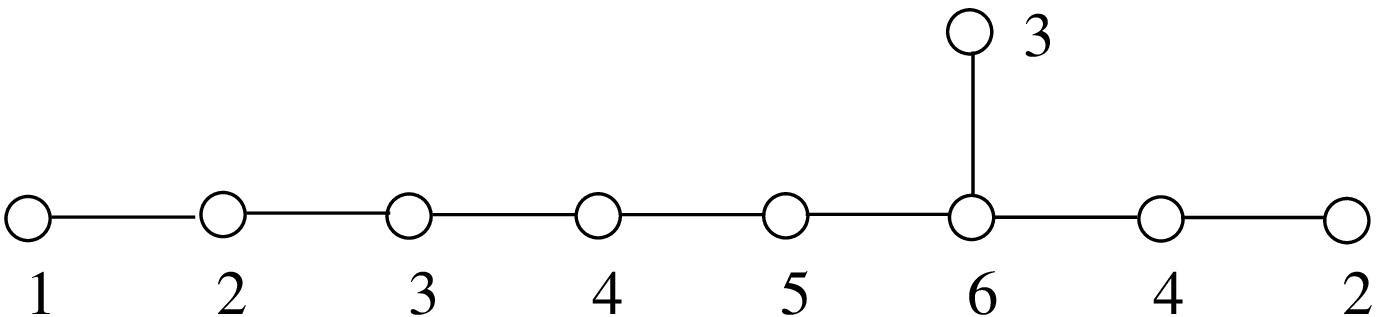}\hskip -0.2cm}
\endinsert}
\vskip -2.5cm\leftskip -2cm
$$\eqalign{{\bf E_8}:\ \ 
&\v^0\ (y^2+x^3+z^6+xyz)\ +\mquad\cr
&\v\ (z^5)\ +\ \v^2\ (z^4+x^2)\ +\ \v^3\ (z^3+y)\ +\cr
&\v^4\ (z^2+x)\ +\ \v^5\ z\ + \ \v^6
}$$
}

\footatend\vfill\supereject\immediate\closeout\rfile\writestoppt
\baselineskip=14pt\centerline{{\bf References}}\bigskip{\frenchspacing%
\parindent=20pt\escapechar=` \input refs.tmp\vfill\eject}\nonfrenchspacing
\end